\documentclass{emulateapj}

\newcommand{\myemail}{kusakabe@astron.s.u-tokyo.ac.jp}

\newcommand{\fesc}{$f_{\rm esc}^{{\rm Ly}\alpha}$}
\newcommand{\fescuv}{$f_{\rm esc}^{\rm UV}$}
\newcommand{\lir}{$L_{\rm TIR}$}
\newcommand{\lirs}{$L_{\rm TIR}^{3\sigma}$}
\newcommand{\luv}{$L_{\rm UV}$}
\newcommand{\sfrir}{$SFR_{\rm IR}$}
\newcommand{\sfruv}{$SFR_{\rm UV}$}
\newcommand{\sfruvc}{$SFR_{\rm UV,corr}$}
\newcommand{\sfrhac}{$SFR_{\rm H\alpha,corr}$}
\newcommand{\sfrtot}{$SFR_{\rm tot,IR+UV}$}
\newcommand{\A}{$A_{1600}$}
\newcommand{\lya}{${\rm Ly}\alpha \xspace$}

\shorttitle{Dust emission of LAEs at z$\sim$2}
\shortauthors{Kusakabe, H. et al.}

\slugcomment{Accepted to ApJL (January 31st, 2015)}

\usepackage{graphicx}
\usepackage{epsfig}
\DeclareGraphicsExtensions{.pdf,.png,.jpg,.eps} 
\usepackage{amssymb, amsmath} 
\usepackage[hyperindex,breaklinks]{hyperref}
\usepackage{xspace}

\begin{document}
%
\title{First Infrared-based implications for the dust attenuation  and star formation of 
typical Ly$\alpha$ emitters \footnotemark[*,**]}


%
\author{Haruka Kusakabe\altaffilmark{1,$\star$}, 
Kazuhiro Shimasaku\altaffilmark{1,2}, 
Kimihiko Nakajima\altaffilmark{3}, 
and Masami Ouchi\altaffilmark{4,5}
}
\altaffiltext{$\star$}{\myemail}
\altaffiltext{1}{Department of Astronomy, Graduate School of Science, The University of Tokyo, 7-3-1 Hongo, Bunkyo-ku, Tokyo 113-0033, Japan}
\altaffiltext{2}{Research Center for the Early Universe, The University of Tokyo, 7-3-1 Hongo, Bunkyo-ku, Tokyo 113-0033, Japan}
\altaffiltext{3}{Observatoire de Gen$\grave{e}$ve, Universit$\grave{e}$ de Gen$\grave{e}$ve, 51 Ch. des Maillettes, 1290 Versoix, Switzerland}
\altaffiltext{4}{Institute for Cosmic Ray Research, The University of Tokyo, 5-1-5 Kashiwanoha, Kashiwa, Chiba 277-8582, Japan}
\altaffiltext{5}{Kavli Institute for the Physics and Mathematics of the Universe (Kavli IPMU, WPI), The University of Tokyo, 5-1-5 Kashiwanoha, Kashiwa, Chiba 277-8583, Japan}
\footnotetext[*]{Based in part on observations made with the Spitzer Space Telescope, which is operated by the Jet Propulsion Laboratory, California Institute of Technology, under a contract with NASA.
}
\footnotetext[**]{Based in part on observations with Herschel, an ESA space observatory with science instruments provided by European-led Principal Investigator consortia and with important participation from NASA.}

%
%
\begin{abstract}

By stacking publicly available deep Spitzer/MIPS 24~$\mu$m and Herschel/PACS images for 213 $z \simeq 2.18$ \lya{} Emitters (LAEs) in GOODS-South, we obtain a strong upper limit to the IR luminosity of typical LAEs and discuss their attenuation curve for the first time. The $3\sigma$ upper limit \lirs $= 1.1 \times 10^{10} L_\odot$, determined from the MIPS data providing the lowest limit, gives $IRX \equiv L_{\rm TIR}/L_{\rm UV} \leq 2.2$. Here we assume that the local calibration between the 8~$\mu$m emission and the dust SED shape and metallicity applies at high redshifts and that our LAEs have low metallicities as suggested by previous studies.
The inferred escape fractions of \lya, $16$--$37\%$, and UV continuum, $\ge 44\%$, are higher than the cosmic averages at the same epoch. The SMC attenuation curve is consistent with the IRX and the UV slope $\beta = -1.4^{+0.2}_{-0.2}$ of our stacked LAE, while the Meurer's relation (Calzetti curve) predicts a 3.8 times higher IRX; we also discuss the validity of PACS-based \lirs{} allowing the Meurer's relation. 
 SED fitting using the Calzetti curve also gives a $\sim 10$ times higher SFR than from the \lirs{} and \luv{}. With $M_{\star}=6.3^{+0.8}_{-2.0} \times10^8 \mathrm{M_{\odot}}$, our LAEs lie on a lower-mass extrapolation of the star formation main sequence at $z \sim 2$, suggesting that the majority of $z \sim 2$ LAEs are mildly star forming with relatively old ages of $\sim 200$ Myr. The faint \lirs{} implies that LAEs contribute little to the faint ($\gtrsim 100~\mu$Jy) submm number counts by ALMA.
\end{abstract}

%
%
\section{Introduction}
\label{IntroductionSection}

The IR luminosity of galaxies combined with the FUV luminosity provides a reliable measure of dust attenuation which is key to obtaining stellar population parameters such as star formation rate and stellar mass. However, at high redshift, because of the limited sensitivities of existing IR telescopes, luminosities have been measured only for relatively luminous galaxies (\lir $\gtrsim 10^{11} L_\odot$ at $z\gtrsim2$, see, e.g., \citet{Elbaz2011,Magnelli2013}) except for a small number of less luminous, lensed galaxies \citep[e.g.,][]{Sklias2014a} and for stacked objects \citep[e.g.,][]{Reddy2012a}. Considering the hierarchical nature of galaxy evolution, we also need to study faint galaxies.

\lya{} emitters (LAEs) are suitable targets for studies of dust emission of faint galaxies because they typically have low ($\lesssim 10^9 M_\odot$) stellar masses \citep[e.g.,][]{Guaita2011, Ono2010b, Nakajima2012}. However, previous studies have failed to individually detect dust emission from LAEs except for rare objects with (U)LIRG-like luminosities \citep{Pentericci2010, Oteo2012}. \citet{Wardlow2014} have stacked Herschel/SPIRE 250--500$\mu$m and LABOCA 870~$\mu$m images of typical LAEs at $z\gtrsim2.8$ to obtain $3\sigma$ upper limits of \lir $\sim 2$--$3 \times 10^{11} L_\odot$, but they have not discussed the amount of dust attenuation. Although, usually, SED fitting has thus been used to derive \A{} for LAEs, \A{} values from SED fitting are sensitively dependent on the attenuation  curve assumed. All of the previous studies of LAEs have assumed the Calzetti curve \citep{Calzetti2000} and obtained a relatively wide range of \A $\lesssim 3$ magnitude (see Table 2 of \citet{Vargas2014}). However, recent observations \citep[e.g.,][]{Reddy2006a, Nordon2013} have found that some high-redshift galaxies favor the Small Magellanic Cloud (SMC) curve \citep{Pettini1998}, which may question the use of the Calzetti curve for LAEs.

In this letter, we use a large sample of $z \simeq 2.18$ LAEs to obtain a strong upper limit to \lir{} using a stacking analysis of deep FIR data in GOODS South. After deriving the \A{} and total SFR, we discuss their attenuation  curve and  the mode of star formation. Also calculated are the submm flux density and the \lya{} and UV escape fractions. We adopt a cosmology with $\Omega_\Lambda$ = 0.7, $\Omega_m$ = 0.3, and $H_{0}$ = 70 km s$^{-1}$ Mpc$^{-1}$ and a Salpeter IMF. 

%
%
%

\section{Data and Sample Selection}

\label{SampleSection}
\label{data}

We select LAEs at $z=2.18\pm0.04$ in the area covered by the deep Spitzer$/$MIPS 24$~\mu$m data from the GOODS survey \citep{Magnelli2011} and Herschel$/$PACS data from the PEP \citep{Lutz2011} and GOODS-Herschel \citep{Elbaz2011} surveys \citep{Magnelli2013}. The selection is performed in color-color space using Subaru/Suprime-Cam narrow-band NB387 data \citep{Nakajima2012} combined with VLT/VIMOS $U$-band \citep{Nonino2009} and MPG 2.2m telescope/WFI $B$-band data \citep{Hildebrandt2006} in essentially the same manner as \citet{Nakajima2012}. After removing a small number of interlopers with a spectroscopic redshift outside of the range probed by NB387 ($z=2.14 - 2.22$) ($\sim 6\%$) and Galactic stars ($\sim 0.4\%$) using the GOODS-MUSIC \citep{Santini2009}, the MUSYC \citep{Cardamone2010}, GMASS \citep{Kurk2013}, and X-ray \citep[Chandra 4Ms;][]{Xue2011} catalogs and AGNs ($\sim 2\%$) detected in X-ray or radio \citep[VLA 1.4-GHz
source catalog;][]{Miller2013}, we obtain a large sample of 213 LAEs down to $NB387=26.4$ magnitude (5$\sigma$, 2$''$ diameter aperture) with average $M_{\rm UV}=-18.7$ magnitude and $\mathrm{\sigma}=0.6$ magnitude.

%
\section{Stacking Analysis}
\label{MethodsSection}

\subsection{FIR wavelengths}

Since none of the 213 LAEs has a counterpart in either the MIPS\footnote[6]{http://irsa.ipac.caltech.edu/cgi-bin/Gator/nph-scan?mission=irsa\&submit=Select\&projshort=SPITZER} or PACS catalog \citep{Magnelli2013} within 1$''$ radius, we perform the stacking analysis at the position of the LAEs in the MIPS/$24\mu$m, PACS/70$~\mu$m, 100$~\mu$m, and 160$~\mu$m bands. For MIPS/$24\mu$m, before stacking, we remove sources listed in the MIPS catalog with a similar method to \citet{Reddy2012a} and subtract a large-scale residual background following the procedure given in \citet{Wuyts2008}. For each band, 50$''$~$\times$~50$''$ cut-out images are median stacked. We then perform aperture photometry at the center of each stacked image on a radius of $3''.0$, $3''.2$, $4''.5$, and $7''.4$ for MIPS/24$~\mu$m, PACS/70$~\mu$m, 100$~\mu$m, and 160$~\mu$m, respectively. No significant signal is detected in any of the four stacked images, where the sky noise for each band is estimated from 1000 realizations generated by bootstrap resampling of the 213 objects as done by \citet{Wardlow2014}. Thus, we derive the $3\sigma$ upper limit of the total flux density for each band by multiplying the $3\sigma$ sky noise by an aperture correction factor of 2.87, 2.45, 1.96, and 1.92 for MIPS/24$~\mu$m, PACS/70$~\mu$m, 100$~\mu$m, and 160$~\mu$m, respectively. The resulting 3$\sigma$ total flux densities are 1.4, 56, 81, and 234~$\mu$Jy, respectively.

We calculate the upper limit of the infrared ($3-1000 \mu$m) luminosity, \lir, of the stacked LAE using the MIPS and PACS $3\sigma$ data separately, by scaling the dust SED templates of local galaxies binned over various properties provided by \citet{Ciesla2014}. For the MIPS data, we scale the lowest-metallicity template to the $24\mu$m limit to obtain \lirs $= 1.1 \times 10^{10} L_\odot$.We adopt this template because it gives the highest (i.e., most conservative) \lirs\ among all. For the PACS data, we fit the templates to a combination of the three data points to obtain \lirs $=1.4 \times 10^{11} L_\odot$ as the most conservative limit. We adopt the MIPS-based \lirs\ as the upper limit of our LAEs and use it in the following section.

Below we discuss the possibility that this MIPS-based \lirs\ may be optimistic. The MIPS/$24\mu$m band measures $\simeq 8\mu$m PAH emission at $z\simeq 2.18$ and there is a well-known trend that the relative contribution of PAH emission to \lir\ decreases with decreasing metallicity \citep{Galliano2011}. The lowest-metallicity template has been constructed from galaxies with $12+\log {\rm O/H} \simeq 8.2-8.4$, which is consistent with those of $z \sim 2-3$ LAEs; bright LAEs, $12+\log {\rm O/H} \simeq 8.0-8.8$ \citep{Nakajima2014}, and faint LAEs, $12+\log {\rm O/H} \simeq 8.2\pm0.1$ \citep{Nakajima2012}. Indeed, this template has a relatively high IR8 (=$L_{8-1000 \mu \rm m}/L_8)$ of 6.7, being close to those of $z\sim 2$ UV-selected galaxies, $\simeq 8-9$ \citep{Reddy2012a}, and significantly higher than the typical value of $z<2.5$ star-forming galaxies \citep[IR8~=~4;][]{Elbaz2011}, and of \citet{Chary2001} templates at the $L_8$ of our sample ($IR8~\sim~4$). Starburst galaxies with high IR surface densities can also have high IR8 \citep{Elbaz2011}. However, even with the PACS-based \lirs, the IR surface density of our LAEs is lower than the threshold for the starburst regime, $\simeq 3 \times 10^{10} L_\odot$ kpc$^{-2}$, where $IR8\simeq 8$. A 1 kpc half-light radius is assumed here (see, e.g., \citet{Hagen2014a}).

Thus, it appears to be unlikely that the MIPS-based \lirs\ is significantly underestimating the true value. Note that the PACS-based \lirs{} divided by the $L_8$ of the lowest-metallicity template scaled to the MIPS photometry, gives $IR8\sim90$. However, we should keep in mind that the derivation of the MIPS-based \lirs{} is based on two important assumptions that the local calibration between the 8~$\mu$m emission and the dust SED shape and metallicity applies at high redshifts and that our LAEs have indeed low metallicities as suggested by previous studies. 

\subsection{Optical to MIR wavelengths}

Forty-four percent of our LAEs are within the coverage of deep HST images from the GOODS and CANDELS surveys. Among them, we use 52 objects with uncontaminated IRAC images to obtain an HST F606W-centered, median image-stacked SED from optical to MIR wavelengths, which we assume to represent the entire sample and use for SED fitting. The images used for stacking are: the WFI $B$ \citep{Hildebrandt2006}, the HST/ACS F606W, F775W, F850LP \citep{Giavalisco2004}\footnote[7]{The F450W image is not used because it is contaminated by \lya{} emission at $z\simeq2.18$.}, the HST/WFC3 F125W, F140W, F160W \citep{Brammer2012, Grogin2011, Koekemoer2011b, Skelton2014}, and the Spitzer/IRAC 3.6~$\mu$m, 4.5~$\mu$m, 5.8~$\mu$m, 8.0~$\mu$m \citep{Damen2011}. In photometry, the aperture radius and the aperture correction for the $B$ and IRAC images are determined following the procedure of \citet{Ono2010b}. The aperture radius for the ACS/WFC images is set to $0''.9$ with an aperture correction factor of 1.07 \citep{Skelton2014}. Table 1 shows the total flux densities of the stacked SED, where the errors are a quadratic sum of the photometric error and error in zero point.

%
%
\section{Results and Discussion}
\label{ResultsSection}

\subsection{Infrared Luminosity and Star Formation Rate}

The $3\sigma$ luminosity obtained in Section 3.1, \lirs~$= 1.1 \times 10^{10} L_\odot$, is 200 times lower than the {\lq}knee{\rq} luminosity of the IR luminosity function at $z\sim 2$ \citep{Magnelli2013}, implying that the majority of $z \sim 2$ LAEs have very faint dust emission.

The 3$\sigma$ upper limit of the dust obscured star formation rate (SFR) is calculated to be \sfrir ${\leq}1.8\ \mathrm{M_{\odot}yr^{-1}}$ using the formula devised by \citet{Kennicutt1998}. 

The unobscured SFR derived from the ultraviolet luminosity of the stacked SED, \luv~$=5.3^{+0.2}_{-0.2}~\times~10^{9}~\mathrm{L_{\odot}}~(=L_{\rm UV,typical})$, using the formula devised by \citet{Kennicutt1998}, is \sfruv$ = 1.5^{+0.07}_{-0.07}\ \mathrm{M_{\odot}yr^{-1}}$. Thus, the ratio of obscured to unobscured SFRs is \sfrir~/~\sfruv~${\leq} 1.2$. This constraint is comparable or stronger than those for $z>2$ LAEs obtained by \citet{Wardlow2014}, \sfrir~/~\sfruv${\leq}2$--$14$. The total SFR is $1.5 \leq $\sfrtot$\ \mathrm{M_{\odot}yr^{-1}}\leq3.3$. \label{eq:SFRtot}
 

\subsection{IRX-$\mathrm{\beta}$ and attenuation  Curves}

\begin{figure}[h]
\includegraphics[width=10cm]{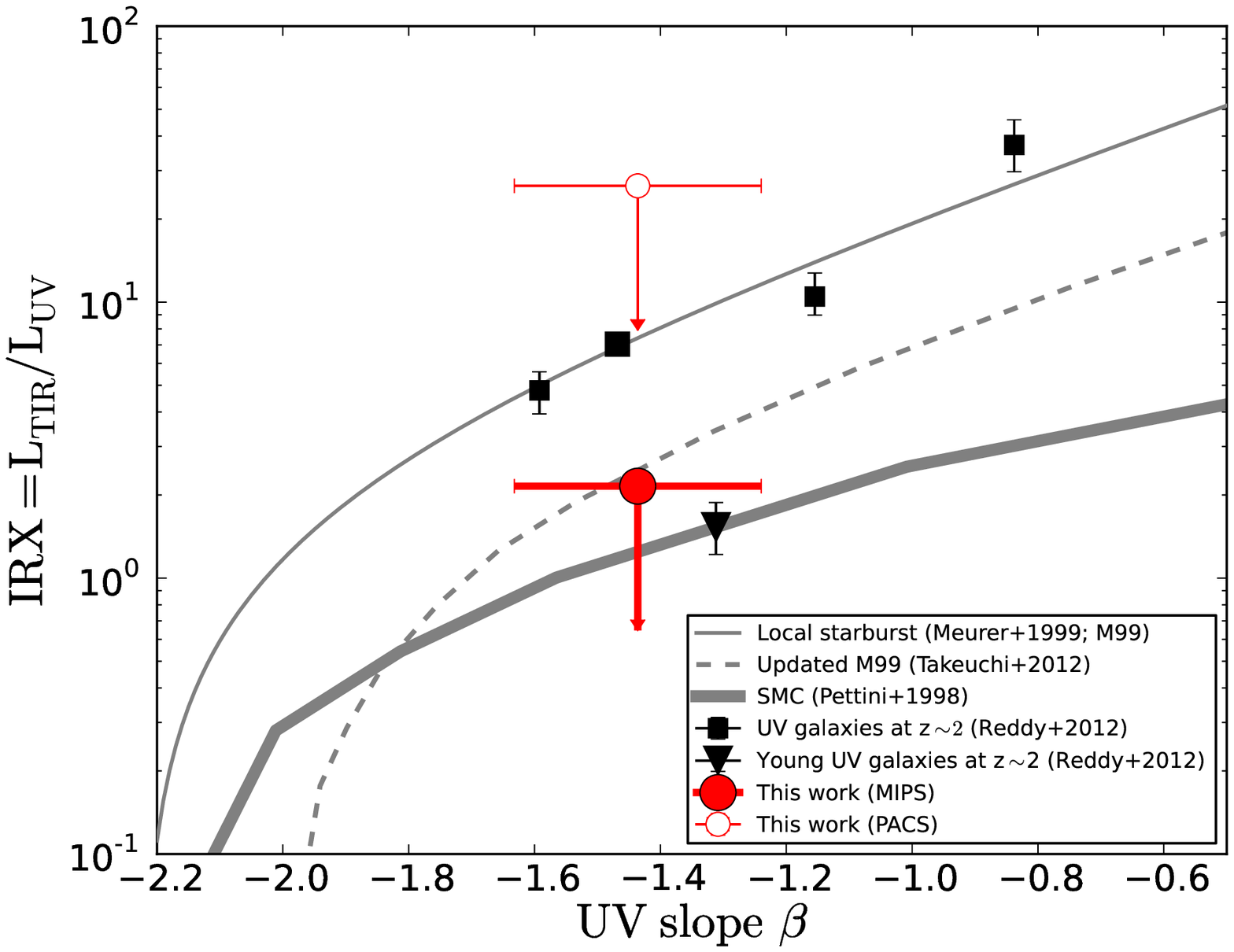}
\vspace{28truemm}
\caption{
The \lir/\luv\ ratio, IRX, plotted against the UV slope, $\beta$.
The red filled and open circles indicate our stacked LAE from the MIPS and PACS data, respectively. 
The black squares are for the stacking results for UV-selected
galaxies binned according to various properties obtained by
\citet{Reddy2012a}
(largest symbol corresponding to the entire sample) 
and the black triangle for the stacking result for 
young galaxies \citep{Reddy2012a}.
Three attenuation  curves are overplotted:
 the local starburst relation (thin solid line; \citet[][M99]{Meurer1999}); 
an updated M99} (dashed line; \citet{Takeuchi2012a});
the SMC curve (thick solid line; \citet{Pettini1998}).

\label{IRX-beta}
\end{figure}

The relation between the IR to UV luminosity ratio, $IRX \equiv$~\lir~/~\luv, and the slope of the UV continuum, $\beta$, is useful for constraining the attenuation  curve of galaxies. Our stacked LAE has $\mathrm{\beta} = -1.4^{+0.2}_{-0.2}$ and $\mathrm{IRX} {\leq}2.2$ using the MIPS/24 $~\mu$m-based \lirs. This IRX corresponds to \A~${\leq}0.9$\ magnitude with the conversion formula by \citet{Overzier2011a}. This low IRX is in accord with the tendency seen in brighter UV-selected galaxies that IRX decreases with decreasing bolometric luminosity \citep[e.g.,][]{Reddy2012a}. As found in Figure \ref{IRX-beta}, these IRX and $\beta$ are inconsistent with the relation for local starburst galaxies \citep{Meurer1999} (M99, almost the same as the Calzetti curve) while consistent with an updated M99 given in \citet{Takeuchi2012a} and the SMC curve \citep{Pettini1998}\footnote[8]{We shift the original relation of \citet{Meurer1999} defined by $L_{\rm FIR}$(40-120$\mu$m) to 0.28dex.}. The original M99 gives a 3.8 times higher IRX at the observed $\beta$. 

All three attenuation  curves assume an intrinsic slope $\beta_{\rm int} \gtrsim -2.2$, while our best-fit SED model spectrum has $\beta_{\rm int} = -2.6$ and $-2.4$ with the M99 and SMC curves, respectively (see sec 4.3). Adopting these bluer $\beta_{\rm int}$ instead of $ \gtrsim -2.2$ increases the inconsistency of the M99 in the IRX--$\beta$ plot while the SMC curve is still consistent.

The $z\sim 2$ UV selected galaxies of \citet{Reddy2012a} are distributed around the M99 except for those with very young ages. LAEs and young UV selected galaxies may have similar attenuation  curves.

The PACS-based \lirs\ gives a high $IRX=28$ falling well above the M99. Indeed, the M99 is allowed in the case of $IRX \ge 8.4$. This happens if the MIPS-based \lirs\ is a factor of $\ge 3.8$ underestimate, i.e. the true IR8 is higher than $26$. However, as discussed in Section 3.1, such a high IR8 appears to be unlikely in our LAEs.



\subsection{SED Fitting}

We perform SED fitting on the stacked SED given in Table 1 to derive stellar population parameters with the procedure given in \citet{Ono2010b}, in which nebular emission is taken into account. We assume constant star formation history following previous SED studies of LAEs and examine the cases of the Calzetti and SMC curves. Although a stacked SED is not necessarily a good representation of individual objects \citep{Vargas2014}, stacking is still useful for our faint objects to obtain an SED covering rest-frame $>1~\mu$m and to consistently compare it with the stacked FIR SED.

\begin{figure*}[h]
\epsscale{0.55}
\plotone{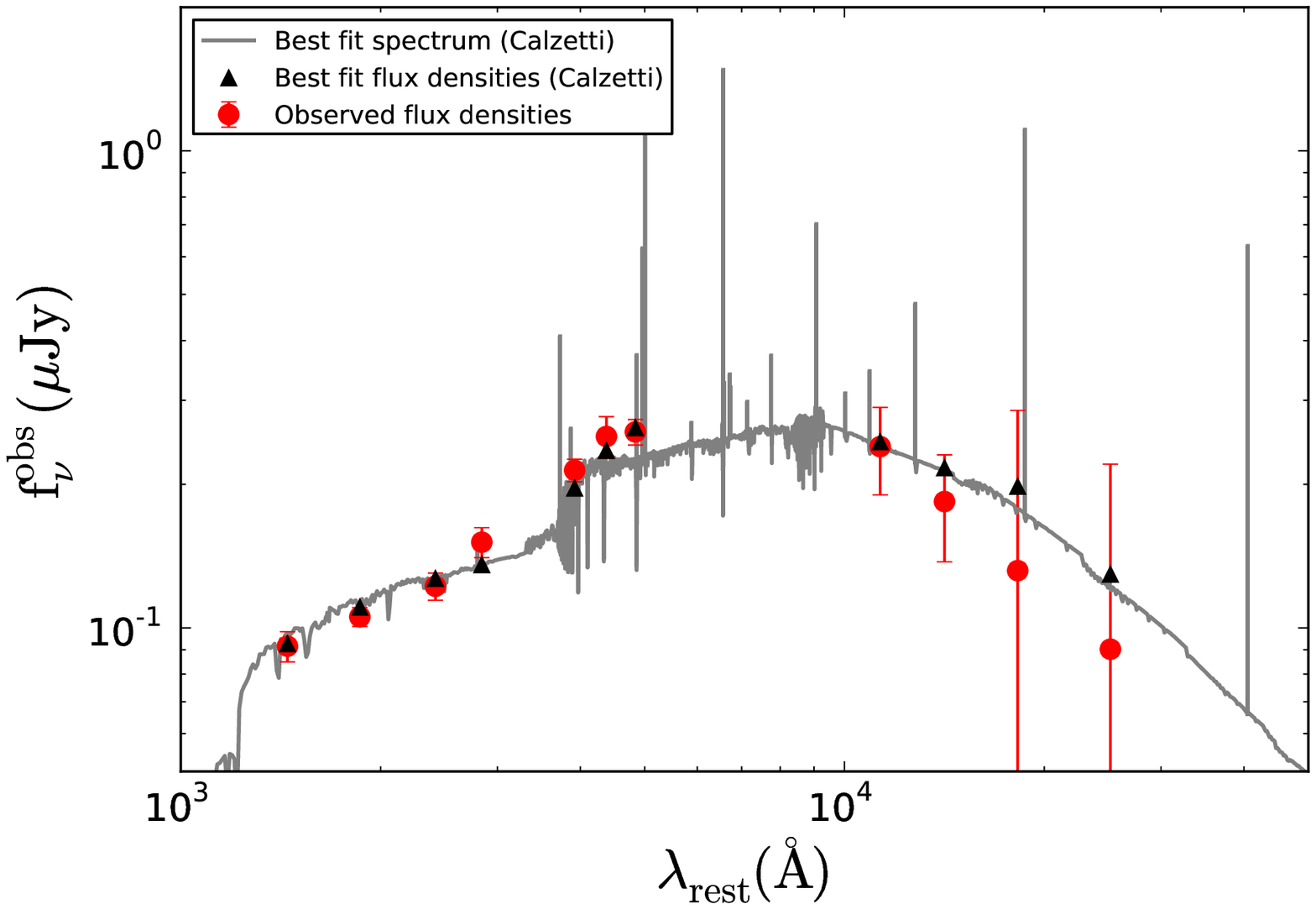}
\plotone{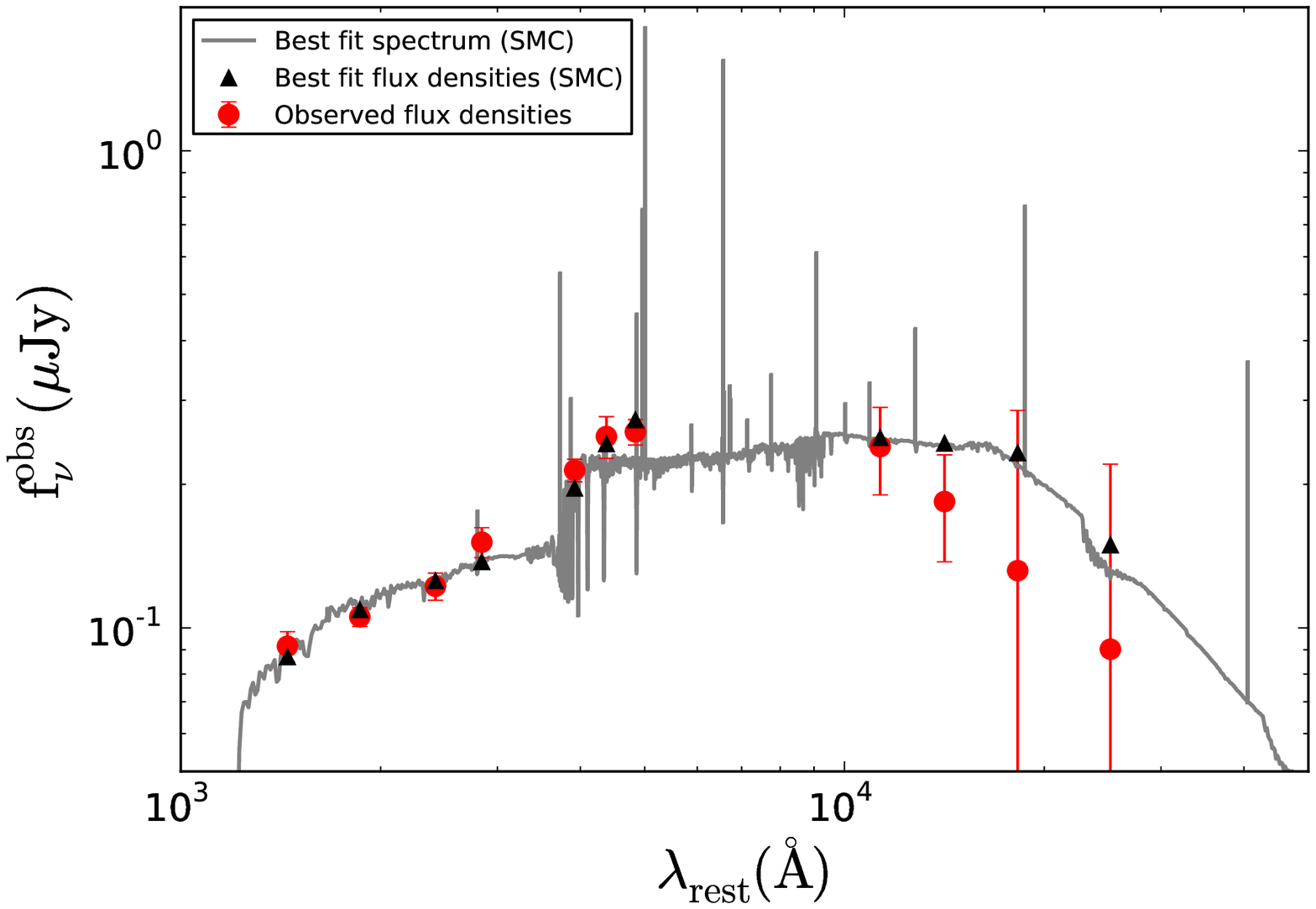}
\vspace{20truemm}
\caption{
Results of SED fitting with the Calzetti curve (left) 
and the SMC curve (right). 
For each panel, the red filled circles show the observed flux 
densities, the gray lines the best-fit model spectrum, 
and the black filled triangles the flux densities calculated 
from the best-fit spectrum.
The two attenuation curves fit the data well.
}
\label{sed_plot}
\end{figure*}

Table \ref{tbl:sed_para} summarizes the best-fit parameters and Figure \ref{sed_plot} compares the best-fit SEDs with the observed one. As expected from the results on the IRX--$\beta$ plot, the fit using the Calzetti curve gives a 2 magnitude higher \A{} and 10 times higher SFR than those calculated from the MIPS-based \lirs, while the results with the SMC curve are roughly consistent. The SFR from the SED fit with the Calzetti curve is significantly larger than even the maximum \sfrtot $=24 M_\odot$ yr$^{-1}$ from the PACS-based \lirs. Both curves give nearly the same stellar mass, because  it is determined essentially from longer wavelengths ($\gtrsim$1~$\mu$m). The Calzetti curve gives an age less than 10 Myr, which is much shorter than dynamical times of LAEs, $\sim60$--$260$ Myr \citep{Rhoads2014}. We also note that the Calzetti curve gives too high an escape fraction of ionizing photons $f_{\rm esc}^{\rm ion}$ compared with observed values for LAEs, $10$--$30\%$ \citep{Nestor2013}. These results suggest that the SMC curve is more appropriate than Calzetti for the majority of LAEs at $z\sim2$. With the SMC curve, our stacked LAE has an relatively old age of 200 Myr. 

\citet{Nakajima2012} have provided a stacked SED of $z\simeq 2.18$ LAEs in the Subaru/XMM-Newton Deep Field for which a less dust-sensitive SFR estimate from narrow-band H$\alpha$ imaging is available. SED fitting to this stacked data finds that the E(B-V) derived assuming the Calzetti curve gives an \sfruvc{} ($\simeq 32\mathrm{M_{\odot}yr^{-1}}$) which is two times higher than the H$\alpha$-based one (\sfrhac{}$\simeq 14 \mathrm{M_{\odot}yr^{-1}}$), while adopting the SMC curve gives a consistent result (\sfruvc{}$\simeq5.7 \mathrm{M_{\odot}yr^{-1}}$, \sfrhac{}$\simeq6.9 \mathrm{M_{\odot}yr^{-1}}$). Notice that $\rm{E(B-V)}_{\rm gas}=\rm{E(B-V)}_{\star}$ is assumed for modest dust-correction of the H$\alpha$ luminosity. Thus we find here that the SMC curve is preferred.


\subsection{Mode of Star Formation}

A relatively tight correlation is seen between SFR and M$_{\star}$ at every redshift \citep{Speagle2014a}. Galaxies on this star formation main sequence (SFMS) are thought to form stars steadily, perhaps being in an equilibrium between gas consumption and accretion, while those above the SFMS are forming stars burstly.

As found in Figure~\ref{MS}, our LAEs lie on a lower-mass extrapolation of the SFMS\footnote[9]{ Figure 17 of the \citet{Daddi2007} suggests that the Calzetti curve applies to BzK galaxies. } 
defined by massive galaxies \citep[$M_\star~>~10^{10}~M_\odot$;][]{Daddi2007, Rodighiero2011}. This suggests that the majority of $z \simeq 2.18$ LAEs are in the {\lq}normal{\rq} star-formation mode and that the SFMS appears to continue down to below $M_\star = 10^{9} M_\odot$. Note that adopting the Calzetti curve leads us to conclude that they are undergoing starburst. On the other hand, most of the extremely luminous, individually detected LAEs at $z \sim 2$ by \citet{Hagen2014a} are burst-like galaxies. The individually detected LAEs by \citet{Vargas2014} also tend to be bursty. Both authors have used the Calzetti curve. One possibility is that the star formation modes of LAEs studied in these papers and our LAEs are different. The other is that some of their LAEs also have an SMC-like attenuation curve and hence their SFRs are predicted to shift towards the SFMS.

\subsection{Can LAEs be Candidates of Faint Submillimeter Galaxies ?}

Recent deep ALMA observations have revealed an abundant population of faint (down to $\gtrsim100~\mu$Jy at $\lambda \simeq 1$ mm) SMGs \citep{Ono2014a}. Optical counterparts to them remain to be identified and LAEs are on a list of potential candidates \citep{Ono2014a}. We find that galaxies with \lirs{} have relatively constant 1.2-mm flux densities of $\sim$~10~$\mu$Jy over a wide redshift range of $z \gtrsim 1$ due to the so-called negative $k$ correction. Therefore, typical LAEs are unlikely to be major counterparts to the faint SMGs unless they significantly evolve in \lir{} with redshift. This is not to rule out the possibility that rare, luminous ( \luv$\gtrsim 10 \times L_{\rm UV,typical} $) LAEs are detectable with ALMA, although a luminous Ly$\alpha$ blob at $z \simeq6.6$, Himiko, is undetected with ALMA at 1.2mm and its strong upper limit, ~52~$\mu$Jy \citep{Ouchi2013a}, is consistent with our value. Deep ALMA observations of luminous LAEs can also be useful, even with non detection, for discriminating between attenuation  curves combined with $\beta$ measurements.

\subsection{Escape Fractions of \lya{} and UV Continuum}
The escape fractions of \lya{} and UV continuum photons are robustly constrained from \lirs. By stacking the 52 LAEs in the NB387 band, we obtain $L({\rm Ly}\alpha) = 5.9^{+0.6}_{-0.6}\times10^{41}$ erg s$^{-1}$, which is converted into $\mathrm{SFR_{Ly\alpha}} = 0.54^{+0.6}_{-0.6}\mathrm{M_{\odot}yr^{-1}}$ \citep{Brocklehurst1971,Kennicutt1998}. Here we have assumed that the NB387 photometry recovers the total \lya{} luminosity.  Since the intrinsic \lya{} luminosity can be calculated from the total SFR, the escape fraction of \lya{} is constrained as: 
\begin{equation}
16\% = \frac{SFR_{\rm Ly\alpha}}{SFR_{\rm UV}+SFR_{\rm IR}^{3\sigma}}
{\le}f_{\rm esc}^{\rm Ly\alpha}
\le\frac{SFR_{\rm Ly\alpha}}{SFR_{\rm UV}} = 37\%.
\end{equation}
This constraint on \fesc{} is roughly consistent with that based on the H$\alpha$ luminosity \citep{Nakajima2012} as well as the lower limits obtained by \citet{Wardlow2014} from FIR stacking. The escape fraction of UV continuum photons is constrained to be:
\begin{equation}
f_{\rm esc}^{\rm UV}
= 10^{-0.4\times{A_{1600}}}
\ge 44\%.
\end{equation}
These \fesc{} and \fescuv{} values are significantly higher than the cosmic averages at $z \sim 2$, \fesc $\simeq 2.8^{+2.6}_{-0.4} \%$ \citep{Hayes2011} and \fescuv $\simeq 22^{+12}_{-7} \%$ \citep{Burgarella2013}, but comparable to those at $z \gtrsim 4$. A number of mechanisms (e.g., outflow and clumpy ISM distribution) have been proposed that make the escape of \lya{} photons easy in LAEs. This study would suggest that for the majority of LAEs absolute low dust attenuation  is an important factor.
\vspace{0.5in}

\begin{figure}[h]
\begin{centering}
\epsscale{1.3}
\plotone{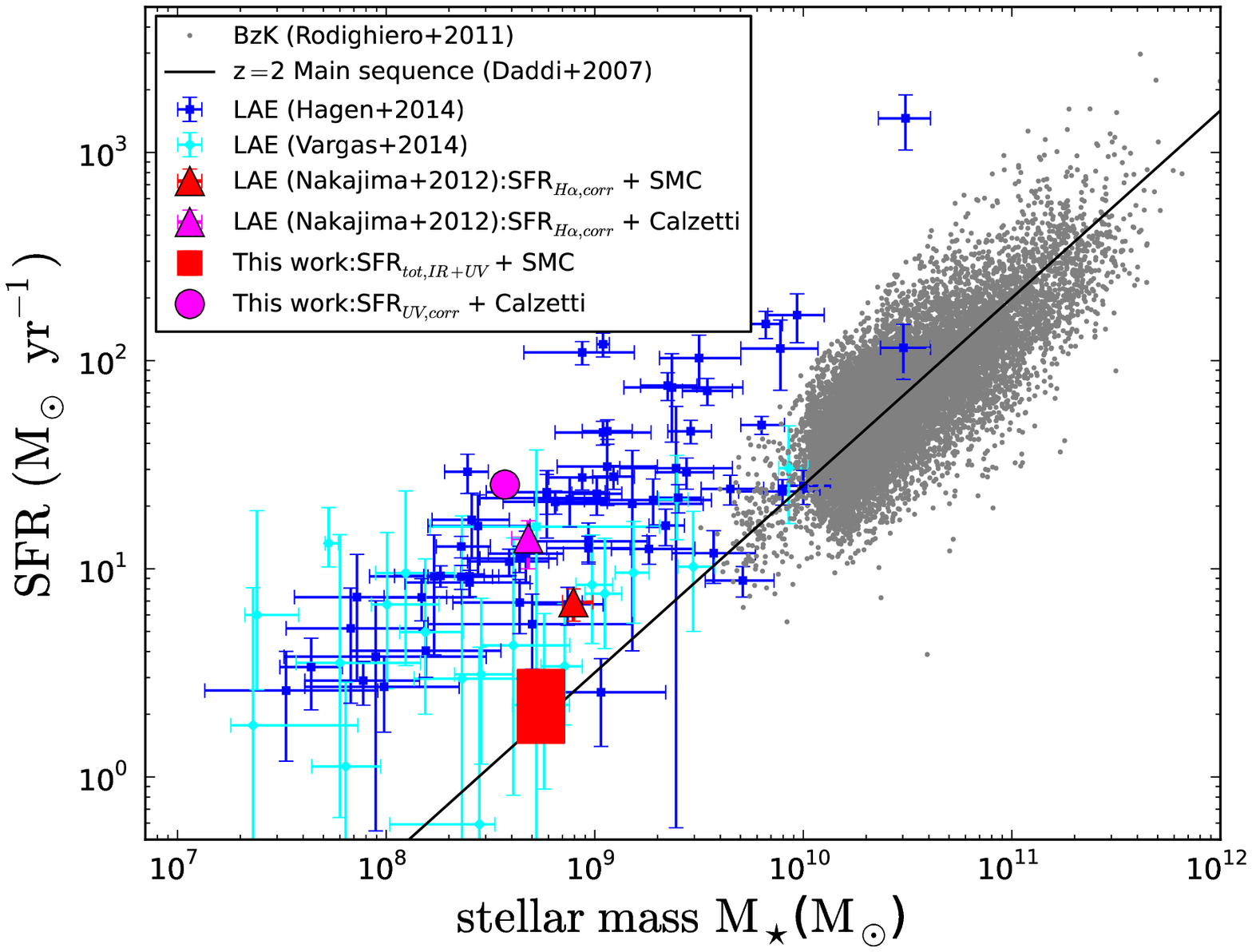}
\vspace{28truemm}
\caption{
$SFR_{\rm tot}$ plotted against M$_{\star}$.
The red rectangular region represents our stacked LAE 
with 
$M_{\star}=6.3^{+0.8}_{-2.0} \mathrm{M_{\odot}}$
from the SMC curve
and \sfrtot$=1.5$--$3.3\ \mathrm{M_{\odot}}$ yr$^{-1}$,
while the magenta filled circle corresponds to the result
from the SED fitting with the Calzetti curve ($SFR=$\sfruvc).
The red and magenta triangles represent LAE from \citet{Nakajima2012}, 
calculated from 
SFR$_{H\alpha,corr}$ using the \A{} obtained 
by the SED fitting with the SMC curve and the Calzetti curve, respectively.
The black line shows the star formation main sequence at $z=2$ 
\citep{Daddi2007} and the gray dots represent 
BzK galaxies \citep{Rodighiero2011}.
The blue and cyan dots are for LAEs 
given in \citet{Hagen2014a} and \citet{Vargas2014}, respectively; 
for both samples the Calzetti curve has been used to derive \A.
}
\label{MS}
\end{centering}
\end{figure}

\acknowledgments
We thank the anonymous referee for helpful comments and suggestions. We are grateful to Yoshiaki Ono for useful comments on SED fitting. We thank Giulia Rodighiero and Alex Hagen for kindly providing their data plotted in Figure \ref{MS}. 
We acknowledge Tamami I. Mori, Tsutomu Takeuchi, Eric Gawiser, Mark Hammonds,  Ronin Wu,  Ryota Kawamata, Takuya Hashimoto, Ryousuke Goto, Shingo Shinogi, and Akisa Ono for useful discussions and comments on this letter.
This work is based on observations taken by the CANDELS Multi-Cycle Treasury Program with the NASA/ESA HST, which is operated by the Association of Universities for Research in Astronomy, Inc., under NASA contract NAS5-26555 and  observations taken by the 3D-HST Treasury Program (GO 12177 and 12328) with the NASA/ESA HST, which is operated by the Association of Universities for Research in Astronomy, Inc., under NASA contract NAS5-26555.
This work is also based on the GOODS-MUSIC catalog.

{\it Facilities: }
\facility{Subaru~(Suprime-Cam)}, %
\facility{Spitzer~(IRAC, MIPS)}, %
\facility{Herschel~ (PACS)}, %
\facility{HST~ (ACS, WFC3)}, %
\facility{VLT~(VIMOS)}, %
\facility{MPG 2.2m telescope~ (WFI)}%

%
%

\begin{deluxetable*}{ccccccccccc}
\tablecolumns{11}
\tabletypesize{\scriptsize}
\tablecaption{Broadband Photometry of the stacked LAE 
from optical to MIR
\label{tbl:photometry_SEDfit}}
\tablewidth{0pt}
\setlength{\tabcolsep}{0.05in}
\tablehead{
\colhead{$B$} &
\colhead{$F606W$} &
\colhead{$F 775W$} &
\colhead{$F850LP$} &
\colhead{$F125W$} &
\colhead{$F140W$} &
\colhead{$F160W$} &
\colhead{$[3.6]$} &
\colhead{$[4.5]$} &
\colhead{$[5.8]$} &
\colhead{$[8.0]$} 
}
\startdata
 $ 0.092$
& $0.11$
& $0.12 $
& $0.15 $
& $0.21$
& $0.25$
& $0.26$
& $0.24$
& $0.18$
& $0.13$
& $0.09$\\
$(0.007)$
& $(0.005)$
& $(0.008)$
& $(0.01)$
& $(0.01)$
& $(0.03)$
& $(0.02)$
& $(0.05)$
& $(0.05 )$
& $(0.2)$
& $(0.1)$
\enddata
\tablecomments{
All flux densities are total flux densities in $\mu$Jy, 
with $1\sigma$ errors shown in parentheses.
}
\end{deluxetable*}


%
%
\begin{deluxetable*}{lccccccc}
\tablecolumns{7}
\tabletypesize{\scriptsize}
\tablecaption{ Results of SED fitting
\label{tbl:sed_para}}
\tablewidth{380pt}
\setlength{\tabcolsep}{0.13in}
\tablehead{
\colhead{attenuation  curve} &
\colhead{$M_{\star}$} &
\colhead{$E(B-V)_{\star}[A_{1600}]$} &
\colhead{Age} &
\colhead{$SFR$} &
\colhead{$f_{\rm esc}^{\rm ion}$} &
\colhead{$\chi^2_r$} &
\colhead{ \sfruvc} \\
\colhead{} &
\colhead{[$10^8 M_{\odot}$]} &
\colhead{[mag]} &
\colhead{[Myr]} &
\colhead{[M$_{\odot}$yr$^{-1}$]} &
\colhead{} &
\colhead{} &
\colhead{[M$_{\odot}$yr$^{-1}$]}
}
\startdata
Calzetti
& $3.7^{+0.1}_{-0.1}$
& $0.3^{+0.00}_{-0.00}[3.0^{+0.0}_{-0.0}]$
& $8.7^{+0.8}_{-1.1}$
& $43^{+4}_{-2}$
& $0.9^{+0.0}_{-0.0}$
& $1.02$ 
& $25^{+1}_{-1}$ \\
\\
SMC 
& $6.3^{+0.8}_{-2.0}$
& $0.10^{+0.02}_{-0.01}[1.2^{+0.2}_{-0.1}]$
& $200^{+50}_{-100}$
& $3.7^{+1.2}_{-0.4}$
& $0.4^{+0.3}_{-0.3}$
& $1.22$
& $4.9^{+1.7}_{-0.3}$
\enddata
\tablecomments{
Metallicity and redshift are fixed to $0.2Z_{\odot}$ and 2.18, 
respectively.
The degree of freedom is $7$.
SFR is not a free parameter in the fit but calculated from 
{$M_{\star}$} and age.
\sfruvc{} is derived from the dust-corrected UV luminosity using \A{}.
}
\end{deluxetable*}


%
%

%
%


%
%


\clearpage
%
%
%
%

\bibliographystyle{mn}                       
\bibliographystyle{apj}                       


%
%

\end{document}